\UseRawInputEncoding
\documentclass[aps,pra,twocolumn,showpacs,groupedaddress]{revtex4}  
\usepackage{graphicx}  
\usepackage{dcolumn}   
\usepackage{bm}        
\usepackage{amssymb}   
\usepackage{amsmath}
\usepackage{float}
\usepackage{ulem}
\usepackage{enumerate}

\begin{document}
				\title{Exotic states induced by co-evolving connection weights and phases in complex networks}
\author{ S. Thamizharasan$^{1}$,  V. K. Chandrasekar$^{2}$,  M. Senthilvelan$^{1}$,  Rico Berner$^{3,4}$, Eckehard Sch\"oll$^3$ and   D. V. Senthilkumar$^5$ }
\address{$^1$ Department of Nonlinear Dynamics, School of Physics, Bharathidasan University, Tiruchirappalli - 620 024, Tamil Nadu, India, \\$^2$Centre for Nonlinear Science \& Engineering, Department of Physics, School of Electrical \& Electronics Engineering, SASTRA Deemed University, Thanjavur -613 401, Tamil Nadu, India. \\
$^3$ Institut f\"ur Theoretische Physik, Technische Universit\"at Berlin, Hardenbergstrasse 36, 10623 Berlin, Germany\\
$^4$ Institut f\"ur Physik, Humboldt-Universit\"at zu Berlin, Newtonstraße 15, 12489 Berlin, Germany\\$^5$School of Physics, Indian Institute of Science Education and Research, Thiruvananthapuram-695 551, Kerala, India.}

\begin{abstract}  	
We consider an adaptive network, whose connection weights co-evolve in congruence with the dynamical states of the local nodes that are under the influence of an external stimulus. The adaptive dynamical system mimics the adaptive synaptic connections common in neuronal networks. The adaptive network under external forcing displays exotic dynamical states such as itinerant chimeras whose population density of coherent and incoherent domains co-evolves with the synaptic connection, bump states and bump frequency cluster states, which do not exist in adaptive networks without forcing. In addition the adaptive network also exhibits partial synchronization patterns such as phase and frequency clusters, forced entrained, and incoherent states.
We introduce two measures for the strength of incoherence based on the standard deviation of the temporally averaged (mean) frequency and on the mean frequency in order to classify the emergent dynamical states as well as their transitions. We provide a two-parameter phase diagram showing the wealth of dynamical states. We additionally deduce the stability condition for the frequency-entrained state. We use the paradigmatic Kuramoto model of phase oscillators which is a simple generic model that has been widely employed in unraveling a plethora of cooperative phenomena in natural and  man-made systems.
\end{abstract}
\pacs{ }
\maketitle

\section{Introduction}
The paradigm of complex networks hosts a plethora of complex collective dynamical states observed in a wide variety of natural and technological systems~\cite{sbvl2006,aaad2008}.
Investigations of the dynamics on and of complex networks have been an active area of research for more than a decade~\cite{tgbb2008,tntv2012,trb2021}.
In particular, adaptive networks are sophisticated complex networks,
a basis for smart systems that self-adapt, in which the connection weights co-evolve along with the dynamical states of the network.
For instance, neurophysiological experiments have revealed that spike timing differences between pre and post synaptic neurons determine the evolution of synaptic connections~\cite{hmjl1997}. Other examples include the reaction rate evolving as a function of the state variable in chemical systems~\cite{sjsk2001}, state-dependent plasticity in epidemics, biological and social systems~\cite{tgcjdd2006,fseo2008,lhck2018}.
Such co-evolution of (synaptic) network connections is thought to provide a basis for higher order brain functions.  Recently, adaptive networks have been intensively investigated and shown to exhibit exotic intriguing dynamical states including chimera states~\cite{tata2009,cbp2011,rgaa2011,lt2014,xzsb2015,llovp2016,dvk2016,scpi2017,dvk2017,vajaa2018,mhm2019,vrrb2019,rbes2019,rbjf2019,rbjs2020}.

In particular, phenomena such as self-organization of hierarchical multilayered structures resulting in multifrequency clusters or chimera states are peculiar to the adaptive networks~\cite{dvk2017,rbes2019}.  
Specifically, a slow adaptation mechanism is identified as pivotal for the emergence of stable multi-clusters. 
A special type of metastable cluster states called "double antipodal"  clusters have been shown to play an important role in achieving the asymptotic dynamics of the adaptive networks~\cite{rbjf2019}. Very recently,  partial synchronization patterns like phase clusters and more complex stable clusters such as splay states, antipodal states and double antipodal states have been found to emerge due to the delicate balance between adaption and multiplexing complex networks, which are otherwise unstable in single layer networks~\cite{rbjs2020}. The results provide exclusive evidence that network adaptation provides a mechanism giving rise to a variety of novel dynamical scenarios that are, however, far from  our current understanding. Therefore, there is  a need for much in-depth investigations.

External stimuli are an inevitable factor to be accounted for along with the adaptive networks as there exist a plethora of phenomena evoked by the external force. For instance, synaptic connections themselves evolve in response to the external stimulus in brain networks~\cite{gpcn1994}. Other examples include circadian rhythms~\cite{mcm1982,jcd2003,rgf2005}, resonances~\cite{izk2002,izk2008,aaz2000,mr2008,eiv2003,asp1997,czjk2001}, event-related desynchronization and synchronization~\cite{gpcn1994}.
Hence, it is very natural to extend the study on 
adaptive networks to systems with external stimulus. Phase diagrams of the forced Kuramoto model have been studied in great detail~\cite{ssyk1986,jaal2005,far2016}.
In this work, we consider an adaptive network of phase oscillators, where the coupling weights co-evolve along with phases, mimicking the spike-timing-dependent synaptic connections between neurons. Under the influence of an external forcing, 
we will show the existence of several exotic self-organizing dynamical states such as conventional chimera states characterized by static coherent and incoherent groups, an only recently discovered type of adaptive chimera termed as {\it itinerant chimera} of two distinct types, characterized by co-evolving coherent and incoherent groups along with the adaptive coupling, {\it bump states} characterized by a localized inactive coherent and coexisting incoherent group, and a bump frequency cluster among other intriguing collective dynamical patterns. We use the Kuramoto order parameter, spatio-temporal plots, instantaneous phase profiles and mean frequency profiles of the oscillators to illustrate the various self-organizing dynamical states. We also introduce the strength of incoherence $S$, based on the standard deviation of the mean frequencies, and $\hat{S}$, based on the mean phase frequencies along with the Kuramoto order parameters to classify the dynamical states and their transitions in  two-parameter phase diagrams. We also deduce the stability condition for the frequency-entrained state.
\begin{figure}[ht!]
		\includegraphics[width=0.500\textwidth]{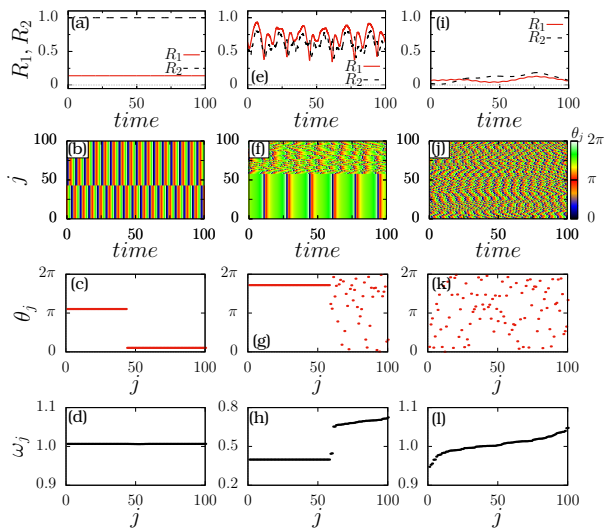}
		\caption{First and second rows depict the order parameters  $R_{(1,2)}(t)$ and space-time plot in the range $t\in(26400, 26500)$, while the third and fourth rows depict a snapshot of 
		instantaneous phases taken at $t=26475$, and temporally averaged mean frequencies,  respectively. The mean is taken over $t\in(24000, 26500)$. (a)-(d) Two-cluster state for $\alpha=0.40$ and  $f=0.0$, 
		(e)-(h) chimera state for $\alpha=0.40$ and $f=0.78$, and (i)-(l) incoherent state for $\alpha=1.54$ and $f=0.20$. Parameters: $\lambda=1$, $\varepsilon=0.005$.}
	\label{fig1}
\end{figure}

The structure of the paper is as follows.  In  section~II, we introduce our model and discuss its constituents in detail. We unravel the different dynamical states exhibited by the globally coupled identical phase oscillators under the influence of the external forcing in Sec. III. We introduce the strength of incoherence
to quantify and classify the different dynamical states in Sec. IV. Dynamical transitions in one- and two-parameter space are discussed in Sec. V. Finally, we provide a summary and conclusion in Sec. VI.

\section{Model}
We consider a network of globally coupled identical phase oscillators under the influence of an external forcing represented by
\begin{eqnarray}
\dot{\psi_i}&=&\sigma-\frac{1}{N}\sum_{j=1}^N\Gamma_{ij}(\psi_i-\psi_j)+f G(\psi_i, \Omega),
\end{eqnarray}
where, $i = 1,2,3, \ldots, N$,  $\psi_i\in[0,2\pi)$ is the phase of the $i^{th}$  oscillator,  $\sigma$ is the identical natural frequency of the oscillators, $f$ and $\Omega$ are the strength and frequency of the external drive, respectively. 
When a coupled oscillator model with external forcing term is reduced to a phase oscillator model, the forcing term will turn out to
depend on the phases of the oscillators.  The latter form corresponds to the active rotator model that can be regarded as a paradigmatic 
model for neurons with type-1 excitability similar to Morris-Lecar models~\cite{bljgo2004}.
The coupling function is given by $\Gamma_{ij}(\psi)=k_{ij}\sin(\psi+\alpha)$, where $k_{ij}$  is the coupling weight from $j^{th}$ to $i^{th}$ oscillator and  $\alpha$ quantifies the phase-lag induced by a small coupling delay~\cite{mmav2018}.
It is to be noted that $\alpha$ maybe also seen as the manifestation of the nonisochronocity parameter of the Stuart-Landau oscillator in the reduced phase model~\cite{jkbook}. 
The adaptation rule for $k_{ij}$ is given by
\begin{align}
 \dot{k}_{ij}=\varepsilon\Lambda_{ij}(\psi_i-\psi_j),
 \end{align}
  where we restrict the coupling values to the region $\vert k_{ij}(t) \vert \le 1$ and the time scale $\varepsilon^{-1}$ is considered to be very much longer than that of the phase oscillators. We choose $2\pi$-periodic functions for the evolution of $k_{ij}$ and $G(\psi, \Omega)$.  For simplicity, $\Lambda(\psi)$ and $G(\psi, \Omega)$ take the form $\Lambda(\psi)=\sin(\psi)$ and $G(\psi,\Omega)=\sin(\psi-\Omega t)$ with the lowest-order Fourier mode.  

Note that the  evolution of the  coupling strength has been chosen as  
\begin{align}
\dot{k}_{ij}=\varepsilon\sin(\psi_i-\psi_j+\beta)
\end{align} 
to account for the co-evolution of the coupling weights and the dynamical states of the nodes of the network~\cite{tata2009}.
Specifically,  neurophysiological experiments indicate that the change in the strength of the synaptic coupling between
neurons depends on the relative timing of the pre- and postsynaptic spikes~\cite{gqbmmp1998}.
Hence, it is natural that the dynamics of the coupling weights  depend on the relative timing of the oscillators.
In particular, models of phase oscillators have been successfully used to explain a variety of dynamical mechanisms induced by 
spike-timing dependent plasticity~\cite{llovp2016,vrrb2019}.

Depending on the value of $\beta$, the dynamics of the coupling weights mimics  Hebbian-like function,
spike-timing dependent plasticity-like function and  anti-Hebbian-like function~\cite{tata2009}.  
In particular,  when $\beta=0$,  $\dot{k}_{ij}=\varepsilon\sin(\psi_i-\psi_j)$, this situation is essentially the
same as that in the case of spike-timing dependent plasticity.
We consider the case of $\beta=0$ and investigate the effect of external stimuli in terms of an  external forcing
on the emerging dynamical states. 

A two-cluster state, a coherent state with a fixed phase relation and an incoherent state with frustration have been reported
with the above evolution equation for the coupling weights in the Kuramoto model in the appropriate ranges of $\beta$~\cite{tata2009}.
When the connection weights are fixed and simply provide heterogeneity in the interaction strength between the Kuramoto oscillators without any
specific evolution rule, it was shown that synchronization can be completely  inhibited when the weights are strongly anti-correlated,
otherwise the synchronization transition observed in the standard Kuramoto model has been generalized to this case of heterogeneous coupling
~\cite{ghpdhz2008}.

Introducing $\psi_i=\theta_i+\Omega t$, the globally coupled identical phase oscillators with the external forcing can be represented  in the rotating frame as
\begin{eqnarray}
\dot{\theta_i}&=&\lambda-\frac{1}{N}\sum_{j=1}^Nk_{ij}\sin(\theta_i-\theta_j+\alpha)+f \sin(\theta_i),
\end{eqnarray}
with $\lambda=\sigma-\Omega$. The latter form of the phase oscillator model is also known from studies on active rotators~\cite{ifsy2020}.
Throughout the paper, we consider $N=100$ oscillators, and initial conditions for which $\theta_i$
and  $k_{ij}$ are uniformly distributed in the interval  $[0, 2\pi)$ and $(-1, 1), \forall j$, respectively, $\lambda=1$ and $\varepsilon=0.005$.

\section{Dynamical states}
\begin{figure}[]
		\includegraphics[width=0.500\textwidth]{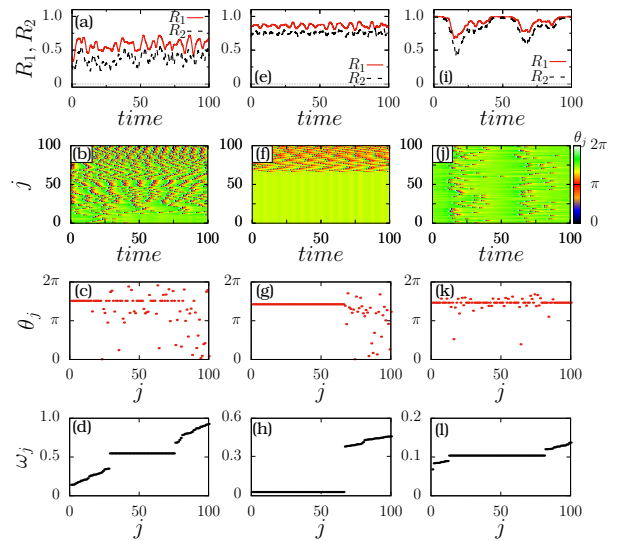}
		\caption{Same plots as in Fig.~\ref{fig1} for different parameters $\alpha$ and $f$.  (a)-(d) itinerant chimera of first kind (IC1) for $\alpha=1.54$ and  $f=0.78$, (e)-(h) bump state $\alpha=0.10$, and $f=0.90$ and (i)-(l) itinerant chimera of second kind  (IC2) for $\alpha=1.54$ and  $f=0.99$.}
			\label{fig3}
\end{figure}

The Kuramoto order parameter 
\begin{eqnarray}
R_l=\left| \frac{1}{N}\sum_{j=1}^N e^{il\theta_j} \right|,~l=1,2 
\end{eqnarray}
characterizes the nature of the dynamical states. $R_1=1$ holds for the completely 
synchronized state, while $R_2=1$ characterizes two cluster (antipodal) states, the latter including the completely synchronized state. The order parameters, spatio-temporal evolution, instantaneous phases and mean frequencies of all oscillators are depicted in the first to fourth row, respectively, of  Figs.~\ref{fig1},~\ref{fig3} and ~\ref{fig5}. 
The unity value of $R_2$ in  Fig.~\ref{fig1}(a)  for $\alpha=0.4$ and $f=0.0$ elucidates that the network exhibits a cluster state, here a two-cluster
 state. However, a small but non-vanishing value of $R_1$ indicates that the two clusters are not of equal size, which is also corroborated by the space-time plot and the snapshot of instantaneous phases in Figs.~\ref{fig1}(b) and ~\ref{fig1}(c), respectively. Note that their mean frequency profile  $\omega_j=\langle \dot{\theta}_j\rangle$, where $\langle \cdot\rangle$ represents a long time average, depicted in Fig.~\ref{fig1}(d), shows that the oscillators are frequency entrained. The instantaneous phases are obtained at $t=26475$ in all the figures of the manuscript unless otherwise specified.

\begin{figure}[]
		\includegraphics[width=0.500\textwidth]{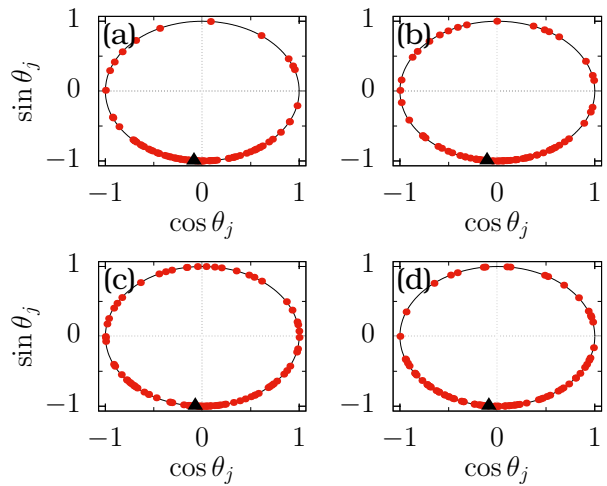}
		\caption{Phase snapshots for an itinerant chimera of the first kind (IC1) at different instants of time. (a) $t=24004$, (b) $t=24007$, (c) $t=24013$ and (d) $t=24034$. The parameter values are the same as in Fig.~\ref{fig3} for IC1.}
	\label{ic1pp}
\end{figure}
The network of globally coupled phase oscillators also exhibits a chimera state  upon increasing the forcing amplitude to $f=0.78$ for the same $\alpha$ (see the middle column of Fig.~\ref{fig1}). The order parameters $R_{(1,2)}$ in Fig.~\ref{fig1}(e) vary within a small range, which is attributed to the coexistence of coherent and incoherent domains, characterizing the chimera state. Distributed frequencies among the oscillators constituting the incoherent domain result in the time-dependent variations of the order parameters.
The coherent and incoherent domains of phase oscillators are well visible in the space-time plot and 
the snapshot of the instantaneous phases depicted in Figs.~\ref{fig1}(f) and ~\ref{fig1}(g), respectively.
The  mean frequency profile (see Fig.~\ref{fig1}(h)) showing distinct groups of phase oscillators with distributed and entrained frequencies is yet another evidence for the existence of chimera in the network of phase oscillators.  In addition, we have further analyzed the emergence of chimera states with respect to the number of coupled  oscillators $N$. Remarkably, we observe chimera states even for a small number of oscillators, see Appendix C.

Random evolution of the instantaneous phase of the oscillators is observed for $\alpha=1.54$ and $f=0.20$. Fluctuating values of $R_{(1,2)}$ near zero in Fig.~\ref{fig1}(i) confirm the incoherent nature of the phase evolution. Further, the space-time plot and the snapshot of the instantaneous phases in Figs.~\ref{fig1}(j) and ~\ref{fig1}(k), respectively, corroborate the random nature of the phase oscillators. Randomness is also observed in the mean frequency profile (see Fig.~\ref{fig1}(l)) corroborating the chaotic distribution of their mean frequencies.

The chimera states reported so far in the literature are mostly characterized by coexisting coherent and incoherent domains with constant population density.
However, turbulent and rotating chimeras are characterized by irregular and periodic temporal  evolution of the coherent domains~\cite{fpkswh2016,oeoek2019}.
In this work, we show the emergence of a recently discovered and thus rather less understood type of chimera characterized by coexisting coherent and incoherent groups, whose population densities co-evolve with the connection weights and the phase of the oscillators.

In general, a chimera state refers to a co-existing group of coherent and incoherent dynamical states arising out of an ensemble of identical
systems.  Various types of chimera states including frequency chimera, amplitude chimera, spiral chimera, traveling chimera,
breathing chimera,  etc., have been reported in the literature~\cite{fpsj2021,oeoek2019}.  The coherent and incoherent groups characterizing 
the chimera state can be  qualitatively identified from the snapshot of the instantaneous phases of the oscillators, snapshot of
the mean frequency and even from their spatiotemporal patterns.  Qualitative measures such as local order parameter, 
strength of incoherence have also been in use to identify and characterize the various chimera states. More 
discussion on the specific types of chimera states, their physical relevance, experimental realizations and their characterizations can be
found in the recent reviews on chimera states~\cite{fpsj2021,smbkb2019,es2020}.

\begin{figure}[]
		\includegraphics[width=0.500\textwidth]{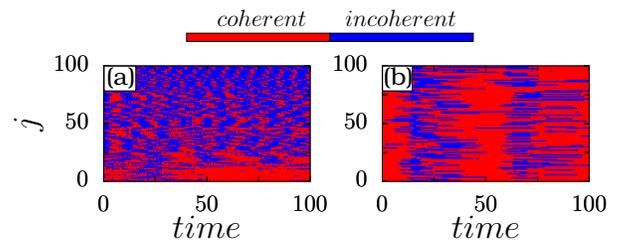}
		\caption{Evolution of the coherent and incoherent groups of the itinerant chimera. Red (light gray) corresponds to the coherent and blue (dark gray) to the incoherent groups. The parameter values are the same as in Fig.~\ref{fig3}.  (a) itinerant chimera of the first kind and (b) itinerant chimera of the second kind.}
	\label{syc}
\end{figure}
The small variations of the order parameters  $R_{1}$ and $R_{2}$ in Fig.~\ref{fig3}(a) for $f=0.78$ indicate the coexistence of coherent and incoherent domains characterizing a chimera state. Phases evolving in concurrence with the adaptive coupling are evident from the space-time plot in Fig.~\ref{fig3}(b). Evolution of connection weights is included in the appendix. Randomly distributed phases without any phase correlation may be observed from the snapshot of the instantaneous phases in Fig.~\ref{fig3}(c). 
Nevertheless, there exists a set of oscillators with nearly identical phases at any given point of time constituting the coherent group while the rest of the oscillators with a drift in their phases constitute the incoherent group which can be best observed for a suitable sorting of the indices $i=1,\ldots, N$ as illustrated in Fig.~\ref{fig3}(b) and ~\ref{fig3}(c). The sorting is chosen according to increasing mean frequencies (see Fig.~~\ref{fig3}(d).)
It is to be noted that the oscillators spontaneously switch between the coherent and incoherent domains as the oscillators self-organize in accordance with the adaptive coupling. Such a self-organizing pattern is reported as ``traveling of the oscillators from one domain to another or as traveling of the chimera core across the network"' and is referred to as ``itinerant chimera"~\cite{dvkvvk2019}.  Since the dynamical nature of the
self-organizing patterns in the first column of Fig.~\ref{fig3} exactly resembles that of the itinerant chimera, we call it an itinerant chimera of first kind.  The existence of coherent and incoherent domains of the itinerant chimera of first kind is also evident from the mean frequency profile depicted in Fig.~\ref{fig3}(d).

\begin{figure}[]
		\includegraphics[width=0.500\textwidth]{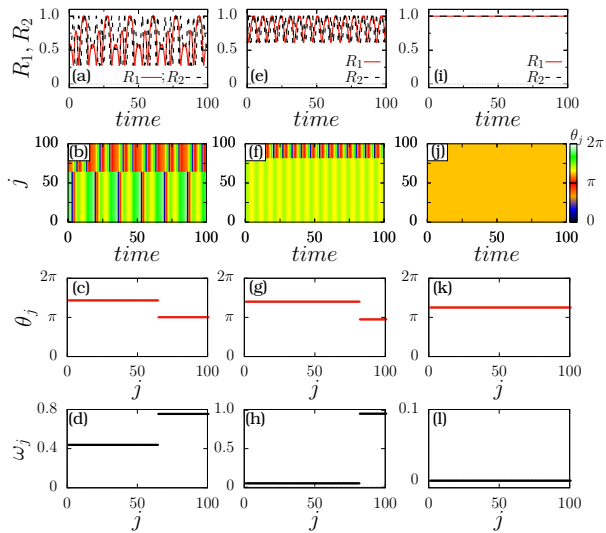}
		\caption{Same characteristic plots as in Fig.~\ref{fig1} for different parameters $\alpha$ and $f$.  (a)-(d) frequency cluster state for $\alpha=0.40$ and  $f=0.83$, (e)-(h) frequency cluster state with bump frequency cluster state  $\alpha=0.60$ and  $f=0.95$  and (i)-(l) forced entrainment state for $\alpha=0.40$ and  $f=1.40$.}
	\label{fig5}
\end{figure}
Phase snapshots of the itinerant chimera of first kind  at four different instants are illustrated in Figs.~\ref{ic1pp}(a)-\ref{ic1pp}(d). The oscillators constituting the coherent domain are depicted as a filled triangle, while the oscillators constituting the incoherent domain are indicated by filled circles. 
Further, the temporal evolution of the oscillators constituting coherent (marked in red/light gray) and incoherent (marked in blue/dark gray) groups of 
the itinerant chimera of the first kind is 
depicted in Fig.~\ref{syc}(a), which resembles the characteristics of the itinerant chimera reported by Kasatkin et al~\cite{dvkvvk2019}.

We observe a bump state upon increasing the forcing to $f=0.9$, where a majority of the oscillators are entrained to the external forcing for $\alpha=0.1$, while the rest evolve independently displaying an active state.  The coexistence of a coherent quiescent state and an incoherent desynchronized oscillating state
is known as a bump state~\cite{mocl2007,crloeo2020,hsda2020,ifoeo2021}. The corresponding order parameters
are depicted in Fig.~\ref{fig3}(e), which oscillate near unity as a majority of the oscillators are locked to the external forcing. The spatio-temporal evolution of this dynamical state with coherent domain constituted by the oscillators entrained with the external forcing and incoherent domain
constituted by independently evolving  phase oscillators is shown in  Fig.~\ref{fig3}(f).  The snapshot of the instantaneous phases (see  Fig.~\ref{fig3}(g)) and the mean frequency profile (see  Fig.~\ref{fig3}(h)) also corroborate  the coexistence of two distinct domains. Note that the phase oscillators with near zero mean frequency and the oscillators with a finite value of average frequencies together constitute the bump state. 
Let $\theta_i=\theta^s$, $i=1,2,\ldots, N_s$  be the oscillators locked to the external force and $\theta_i=\theta_i^d$,  $i=N_s+1,N_s+2,\ldots, N$
be the oscillators evolving independently, then the instantaneous angular velocity corresponding to the coherent quiescent state and the  incoherent desynchronized oscillating state
can be written as
 \begin{widetext}
 \begin{subequations}
\begin{align}
\dot{\theta^s} &= \lambda-\frac{N_s}{N}k_1\sin \alpha-\frac{1}{N}\sum_{j=N_s+1}^N k_{sj}\sin(\theta^s-\theta_j^d+\alpha)+f\sin\theta^s,  \\
\dot{\theta_i^d} &= \lambda-\frac{N_s}{N}k_2\sin (\theta_i^d-\theta^s+\alpha)-\frac{1}{N}\sum_{j=N_s+1}^Nk_{ij}\sin(\theta_i^d-\theta_j^d+\alpha)+f\sin\theta_i^d,
\end{align}
\label{eq:ps}
 \end{subequations}
\end{widetext}
where $k_1=\frac{1}{N_s}\sum_{j=1}^{N_s} k_{ij},  ~\forall ~i=\{1,2,\ldots, N_s\}$ and $k_2=\frac{1}{N_s}\sum_{j=1}^{N_s} k_{ij}, ~\forall~i =\{N_{s}+1,N_s+2,\ldots, N\}$. 
 Indeed, we have confirmed from the numerical analysis that the entrained  oscillators as well as those evolving independently satisfy the above equations.

\begin{figure}[]
		\includegraphics[width=0.500\textwidth]{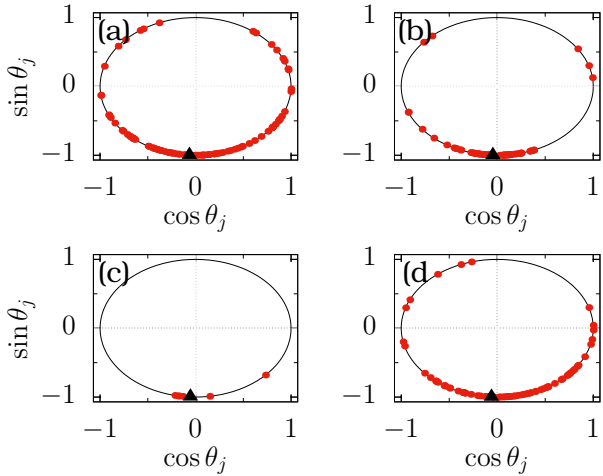}
		\caption{Phase snapshots of the itinerant chimera of the second kind at different instants of time. 
		  (a) $t=24040$, (b) $t=24055$, (c) $t=24280$ and (d) $t=24317$.  The parameter values are the same as in Fig.~\ref{fig3} for the itinerant chimera of the second kind.}
	\label{ic2pp}
\end{figure}

The itinerant chimera of the second kind is depicted in the third column of Fig.~\ref{fig3} for $\alpha=1.54$ and $f=0.99$.   
The essential difference between itinerant chimera of the first and second kind is that the latter is characterized by a high degree of phase-synchronization at certain intervals of time as quantified by the epochs of the order parameters $R_{1,2}$ near unity in Fig.~\ref{fig3}(i). We can speculate that such a chimera state with a large scale phase entrainment may be related to neuropathological states 
with characteristic large-scale synchronization (epochs of abnormal/massive synchronization) such as  in epileptic seizures~\cite{pjuws2006,mfrb2020,es2020}.
Small range variations of the order parameter in Fig.~\ref{fig3}(i) indicate the chimera state.   Relatively high degree of phase coherence coexisting with a low degree of phase coherence, 
compared to that of the itinerant chimera of the first kind, is evident from the space-time plot in Fig.~\ref{fig3}(j) and from the snapshot of the instantaneous phases in Fig.~\ref{fig3}(k). 
The indices of the globally coupled oscillators are reordered suitably to visualize the coherent and incoherent domains clearly according to increasing mean frequency.
It is to be noted that the range of the mean frequency distribution is rather narrow in the range $(0, 0.2)$ for the itinerant chimera of the second kind (see Fig.~\ref{fig3}(l)), whereas that of the itinerant chimera of the first kind is distributed in a much broader range $(0, 1.0)$ corroborating their degree of phase coherence. The plots of phase snapshots of the itinerant chimera of the second kind at four different instants of time are illustrated in Figs.~\ref{ic2pp}(a)-\ref{ic2pp}(d). The oscillators constituting the coherent group are depicted by a filled triangle, while the oscillators constituting the incoherent group are indicated by filled circles. 
 It can be seen from these figures that the 
number of oscillators  constituting coherent and incoherent groups vary as they self-organize in concurrence with the adaptive coupling.
 Further, the temporal evolution of the oscillators constituting coherent (marked in red/light gray)  and incoherent (marked in blue/dark gray) groups of the itinerant chimera of the second kind is depicted in Fig.~\ref{syc}(b).

The network of globally coupled phase oscillators with external forcing exhibits a frequency cluster state for  $\alpha=0.40$ and $f=0.83$. The order parameters $R_1$ and $R_2$ corresponding to the frequency cluster  vary between the values $0.25$ and $1$ (see Fig.~\ref{fig5}(a)).  The oscillators split into two
clusters of different frequency of oscillations (see  Fig.~\ref{fig5}(b)) constituting the frequency cluster state, which is also evident from the  snapshot of the 
instantaneous phases in Fig.~\ref{fig5}(c) and  from the  mean frequency profile depicted in Fig.~\ref{fig5}(d).  
Similar partially synchronized patterns like phase clusters and more complex stable clusters are shown to emerge due to the delicate balance between adaptation and multiplexing in complex networks~\cite{rbjs2020,dvk2016}.
 
In Fig.~\ref{fig5}(e-h), we show another frequency cluster for $\alpha=0.60$ and $f=0.95$. Here, both order parameters $R_1$ and $R_2$ are uniformly oscillating between $0.5$ and $1$ as shown in Fig.~\ref{fig5}(e).  The spatio-temporal plot (see Fig.~\ref{fig5}(f)) elucidates the coexistence of a cluster with fully pronounced spike oscillations and a cluster with small amplitude sub-threshold oscillations induced by the other cluster, which we call the bump frequency cluster state. The instantaneous phases (Fig.~\ref{fig5}(g)) and the mean frequency profile (Fig.~\ref{fig5}(h)) illustrate that 
phases of one group of oscillators are nearly locked with the frequency of the external forcing, while those of the second group undergo a complete cycle indicating the 
bump frequency cluster state.

A forced entrained state is observed for
$\alpha=0.40$ and $f=1.40$, where all the oscillators are entrained by the external forcing. Both order parameters acquire unity values (see Fig.~\ref{fig5}(i)) corroborating the forced entrained state. The oscillator phases acquire the same phase as that of the external forcing  as depicted in the space-time plot in Fig.~\ref{fig5}(j) and in the snapshot of the instantaneous phases in Fig.~\ref{fig5}(k). The  mean frequency profile in Fig.~\ref{fig5}(l) also confirms the frequency entrained state.  

The coupling weights reach the frozen states asymptotically  for two cluster  and forced entrained states (see Appendix A), while for all other states 
the coupling weights evolve with time.  The time variation of  the coupling weights are larger in the chimera, incoherent state, itinerant chimera of first and second kind
but we found very negligible variations in frequency cluster, bump state and bump frequency cluster state.  Further one to one correlation of the dynamical states with its corresponding
evolution of the coupling weight matrices  require more in depth statistical analysis of the latter.
In the following, we will introduce the strength of incoherence~\cite{rgvkc2014} to quantify the different dynamical states and to classify them in one- and two-parameter space plots. 
\begin{figure}[]
		\includegraphics[width=0.500\textwidth]{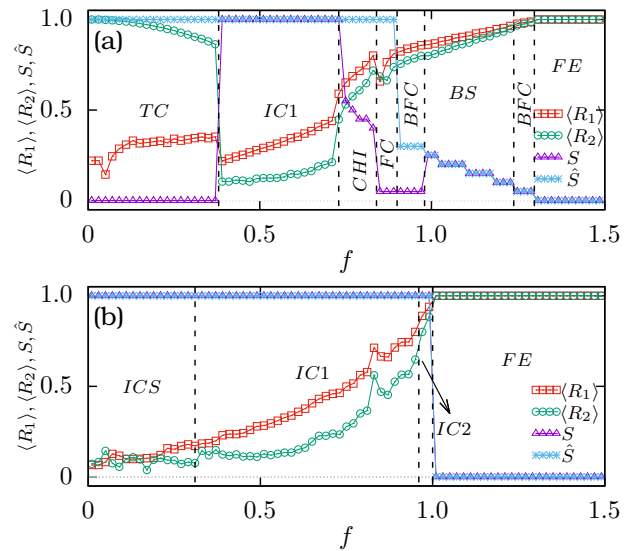}
		\caption{Time-averaged order parameters ($\langle R_1\rangle$ and $\langle R_2\rangle$) and the  strength of incoherence  ($S$ and $\hat{S}$) as a function of the external forcing strength $f$
		elucidating  the dynamical transitions for the parameter values (a) $\alpha=0.30$ and (b) $\alpha=1.54$. Abbreviations: TC - two cluster state, CHI - chimera state, ICS - incoherent state, IC1 -  itinerant chimera states of first and IC2 - second kind, BS - bump state, FC - frequency cluster state, BFC - bump frequency cluster state, FE - forced entertainment state.}
		\label{fig7}
\end{figure}
\section{Strength of incoherence}
In order to define a measure of incoherence, we divide the $N$ oscillators into $M$ (even) bins each of size  $n=N/M$.  We estimate the local standard deviation  ($\sigma_m$) of the frequencies at each bin as
\begin{align}
\sigma_m&=\sqrt{\frac{1}{n}\sum_{j=n(m-1)+1}^{mn}\left[\omega_{j}-\bar{\omega}_m\right]^2},\\ \nonumber
&\qquad\qquad\qquad\qquad\qquad ~m=1, 2, \dots, M.
\label{sd}
\end{align} 
with $\bar{\omega}_m=\frac{1}{n}\sum_{j=n(m-1)+1}^{mn}\omega_{j}$.  Note that $\omega_j$  is the time averaged frequency of the $j$th oscillator.
Now, the strength of incoherence $S$ is calculated using the formula~\cite{rgvkc2014}
\begin{eqnarray}
S=1-\frac{\sum_{m=1}^{M}s_m}{M}, \qquad s_m=\Theta(\delta-\sigma_m),
\label{si}
\end{eqnarray}
where $\Theta$ is the Heaviside step function and $\delta$ is a predefined threshold.  The local standard deviation $\sigma_m$ becomes zero if the network of oscillators is frequency-entrained as in the two-cluster state (see Fig.~\ref{fig1}(a)-\ref{fig1}(c)).  Since  $\sigma_m<\delta$, $s_m$ becomes unity for all $m$.
As a consequence, the strength of incoherence for the phase clusters with frequency entrainment will be  $S=0$. In contrast, for the evolution of phases  with random frequencies such as   the incoherent state (see Fig.~\ref{fig1}(i)-\ref{fig1}(l)), $\sigma_m$ takes a finite value larger than the
 predefined threshold $\delta$  and hence  $s_m=0$  for all $m$ resulting in the strength of incoherence $S=1$ (refer Appendix B for more details). 
 On the other hand, for the case of conventional chimeras,  frequencies  of the phase oscillators that constitute the coherent domain are entrained, whereas   those of the incoherent domain are completely uncorrelated, and hence $S$ takes a value between zero and unity depending on the number of oscillators  constituting the coherent domain. Since the phases evolve with random frequencies  for the itinerant chimeras, $S$ for these states  also takes the value unity similar to the incoherent state.  However,  $S$ for the mean frequency profile of the itinerant chimera of the first kind (see Fig.~\ref{fig3}(d)) and of the second kind (see Fig.~\ref{fig3}(l)), where the oscillator indices are reordered appropriately, takes a value between zero and unity in analogy with the chimera state.
In the bump state, the coherent domain is entrained with the frequency of the
 external forcing while the oscillators of the incoherent domain evolve with random frequencies and hence  the strength of incoherence for the bump state is given by $0<S<1$.  For  the frequency clusters, there is a discontinuous jump in the frequency and  $s_m=0$ only in the bin where such discontinuity exists, while $s_m=1$ in the other bins.  Hence the strength of incoherence  takes the value $1/M$
  close to zero due to frequency entrainment in the $M-1$ bins~\cite{note}.  For a suitable choice of $M$ the value of $S$ might be exactly zero. In any case, $S$ depends on the choice of $M$), whereas in the case of the bump and chimera states $S$ takes a large value  between $0$ and $1$ as there is an appreciable number of bins with a large standard deviation.
The  natural frequencies of the phase oscillators are completely entrained to that of the stimulus in a forced entrained state (see Fig.~\ref{fig5}(g)-\ref{fig5}(i)),
and hence $s_m=1$ in all the  $M$ bins and as a consequence this state is characterized by $S=0$.
 
States that are entrained to the external stimulus such as forced entrained  and bump frequency cluster states   are characterized by near zero mean of the  time-averaged frequencies and can be more clearly distinguished using the mean of the time-averaged frequencies in each bin. In particular, the conventional chimera state, where the strength of the incoherence takes a value $0<S<1$,  can be clearly distinguished using the mean of the time-averaged frequencies in each bin instead of their standard deviation in Eq.~(\ref{si}). For this purpose, we define a modified strength of incoherence using the  mean of the time-averaged frequencies in each bin as
\begin{eqnarray}
\hat{S}=1-\frac{\sum_{m=1}^{M} \hat{s}_m}{M}, \qquad \hat{s}_m=\Theta(\delta-\bar{\omega}_m),
\label{si1}
\end{eqnarray} 
where $\bar{\omega}_m$ in each bin for the two-cluster state has a finite value and hence the corresponding
$\hat{s}_m=0$.  As a consequence $\hat{S}=1$  for the two-cluster state, whereas $S=0$ for this state. For the same reason, $\hat{S}=1$ for the chimera, incoherent, and itinerant chimera of the first and second kind, and forced entrained states. Note that for the forced entrainment states $S=\hat{S}=0$ because of the
complete entrainment with the stimulus.
Bump and the bump frequency cluster states are characterized by intermediate values between $0$ and $1$  for $S$ and $\hat{S}$ as a fraction of the phase oscillators are entrained with the stimulus
while the rest have a finite time-averaged frequency.

In the next section, we classify the dynamical states using the time-averaged order parameters $\langle R_1\rangle$,  $\langle R_2\rangle$, along with the strengths of incoherence $S$ and $\hat{S}$ in one- and two-parameter phases.

\section{Map of regimes}
Transitions in parameter space among the different dynamical states are depicted in Figs.~\ref{fig7}(a) and ~\ref{fig7}(b) as a function of the strength of the external stimulus for 
two different $\alpha=0.3$ and $\alpha=1.54$, respectively. The degree of the strength of incoherence and that of the time-averaged order parameters have been used to clearly distinguish the dynamical states and their transitions. Two-cluster state  (TC) exists in the range of $f \in(0, 0.38)$ (see Fig.~\ref{fig7}(a)) and there is a transition from two-cluster to itinerant chimera states of the first kind (IC1)  which prevails in the range $f \in[0.38, 0.73)$.  Upon increasing $f$ further, there is a transition from IC1 to forced entrained (FE) state via chimera (CHI), frequency cluster (FC), bump frequency cluster (BFC), and  bump states (BS). Chimera states exist in the range $f \in[0.73, 0.84)$, 
frequency cluster states in the range $f \in[0.84, 0.89)$, bump frequency cluster states in the range $f \in[0.89, 0.98)$ and $f \in[1.24, 1.30)$, bump states in the range $f \in[0.98, 1.24)$ and forced entrainment states in the range $f \in[1.30, 1.5)$. The transition from incoherent to forced entrainment states via itinerant chimera states of the first (IC1) and second kind (IC2)  is depicted in  Fig.~\ref{fig7}(b) for $\alpha=1.54$.

\begin{figure}[!ht]
		\includegraphics[width=0.5\textwidth]{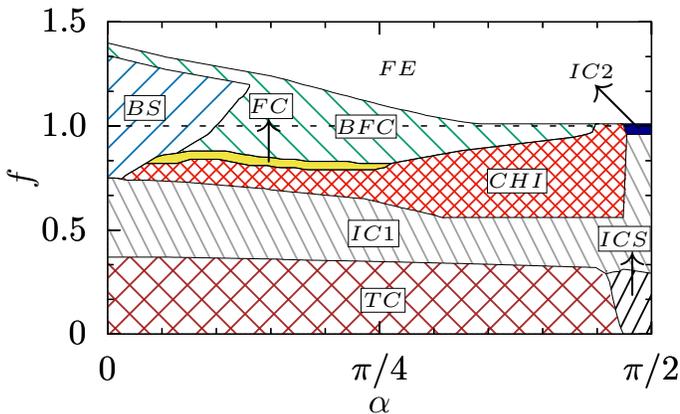}
	\caption{Two-parameter phase diagram in the ($\alpha$, $f$) parameter space depicting various collective dynamical states of the adaptive network with external periodic forcing obtained from one initial condition. The dynamical states in the parameter space indicated by TC, ICS, IC1, CHI, FC, BS, BFC, IC2, FE corresponds to	two-cluster, incoherent state, itinerant chimera of the first kind, chimera, frequency cluster, bump state, bump frequency cluster, itinerant chimera of the second kind and forced entrained state, respectively. The dashed line corresponds to the condition $f\geq \lambda$ with $\lambda=1$ for the forced entrained state.}
	\label{fig8}
\end{figure}
Complementing the analysis in Fig.~\ref{fig7}, the dynamical scenarios in the $(\alpha, f)$  parameter space are shown in  Fig.~\ref{fig8}. Here the initial conditions are distributed such that for all $i, j =1\dots,N$ the coupling weights $k_{ij}$ are uniformly distributed between $-1$ to $1$, while the phases are uniformly distributed between $[0,2\pi)$
  in the entire explored range of $\alpha$ and $f$.  For low values of $f$, that is approximately $f\in(0,0.38)$,  there is a transition
from TC state to incoherent  state (ICS)  as a function of $\alpha$.  An itinerant chimera of first kind  is observed in the entire explored range of $\alpha$ in a narrow range of $f\in(0.38, 0.56)$. Moreover, the itinerant chimera of first kind re-emerges in the range of  $f\in(0.56, 0.96)$  via the chimera state (CHI) for larger values of $\alpha$. The spread of the bump state (BS) increases as a function of $\alpha$ and $f$ until $\alpha=0.42$, which then decreases until $f=1.20$. In a narrow range of $f\in(0.75, 0.79)$, there is a transition from the bump state to an itinerant chimera of first kind via a chimera state for increasing $\alpha$. In the range of $\alpha\in(0.11, 0.81)$ there is a very fine region of frequency cluster states separating  the chimera state and bump frequency cluster state. Then there is a transition from a bump state to an itinerant chimera of first kind via bump frequency and chimera state, which is then followed by a transition from a bump state to an itinerant chimera of second kind via a bump frequency cluster and chimera state as a function of $\alpha$ in the appropriate ranges of $f$. There is also a transition from a bump state to forced entrainment state as a function of $\alpha$ for larger values of $f$. The forced entrainment state prevails in the entire parameter space for $f>1.4$.  The dashed line corresponds to the stability condition of the forced entrained state above which this state is stable, and it can be deduced as shown in the following.

\begin{figure}[!ht]
	\includegraphics[width=0.5\textwidth]{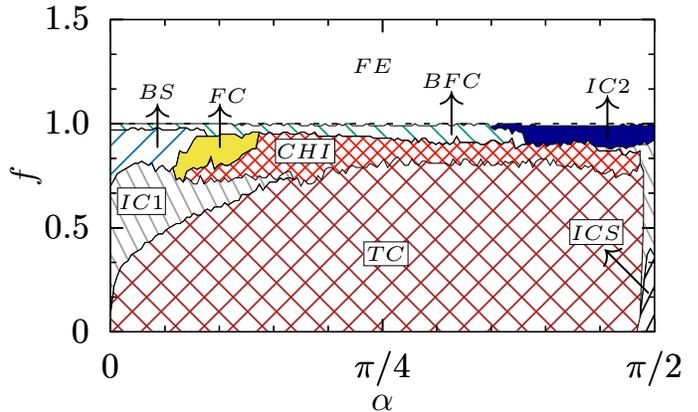}
	\caption{Same as Fig.~\ref{fig8} but for  different initial conditions.}
	\label{fig9}
\end{figure}
In the forced entrained state, all the oscillators are entrained to the same phase $\theta_i=\theta^*$ given by
\begin{eqnarray}
\theta^*=\sin^{-1}\left(\frac{\sum_{j}k_{ij}\sin(\alpha)-N\lambda}{Nf}\right),
\label{eq1}
\end{eqnarray}
which possesses two solutions only if the corresponding coupling weights fulfill the condition that $\frac{1}{N}\sum^N_{j=1}k^*_{ij}=\eta$ is independent of $i$ and further the condition $|\eta\sin(\alpha)/N-\lambda|\le f$ is met. With $\sin ^{-1}:[-1,1]\to [-\pi/2,\pi/2]$ the two solutions are given by $\theta^*_1=\theta^*$ and $\theta^*_2 =\pi-\theta^*$. Note that for the forced entrained state, we have $\dot{k}_{ij}=0$ for all $i,j$. 
The stability of the fixed point can be deduced using linear stability analysis.
The diagonal components ($DF_{ii}$) of the Jacobian matrix $J$  are 
\begin{eqnarray}
DF_{ii}&=&\left(\frac{k_{ii}}{N}-\eta\right)\cos\alpha+f\cos(\theta_{i}^*)\nonumber
\end{eqnarray}
The off-diagonal components ($DF_{ij}$) of the Jacobian matrix $J$ are 
\begin{eqnarray}
DF_{ij}&=& \frac{k^*_{ij}}{N}\cos\alpha. \nonumber
\end{eqnarray}
The characteristic equation of the Jacobian matrix $J$ is given by 
\begin{eqnarray}
(\mu+\eta\cos\alpha-f\cos(\theta^*))^N=0, \nonumber
\end{eqnarray}
 which leads to the N-degenerate eigenvalues
\begin{eqnarray}
\mu_k=-\eta\cos\alpha\pm f\sqrt{1-\bigg(\frac{\eta\sin\alpha -\lambda}{f}\bigg)^2}, \nonumber
\end{eqnarray}
for $k=0,\dots, N-1$ and ``+" and ``-" corresponding to the solutions $\theta^*_1$ and $\theta^*_2$, respectively. 
The forced locked state emerge via a fold bifurcation at 
\begin{eqnarray}
\sqrt{\bigg(\eta\cos\alpha\bigg)^2+\bigg(\eta\sin\alpha -\lambda\bigg)^2}=\pm  f.
\end{eqnarray}

Note that for a suitable choice of initial conditions one may observe $\frac{1}{N}\sum_{j}k^*_{ij}=\eta=0$. Hence, the condition for the existence of a stable forced entrained state simplifies to $|\lambda|\le f$. 
To show the validity of the derived condition, the dynamical scenarios in the $(\alpha, f)$ parameter space for uniformly distributed initial conditions that lead to $\eta=0$ are explored and the results are depicted in Fig.~\ref{fig9}. The dynamical transitions are almost similar to those in Fig.~\ref{fig8} except for an increase or decrease in the spread of the observed  dynamical states.  It is to be noted that the stability condition  given by $f\geq \vert\lambda\vert$ with $\lambda=1$ (indicated by the dashed line),  above which the forced entrained state is stable, exactly matches with the simulation results for the chosen set of initial conditions, in accordance with the condition derived from the stability analysis.

\begin{figure}[!ht]
		\includegraphics[width=0.5\textwidth]{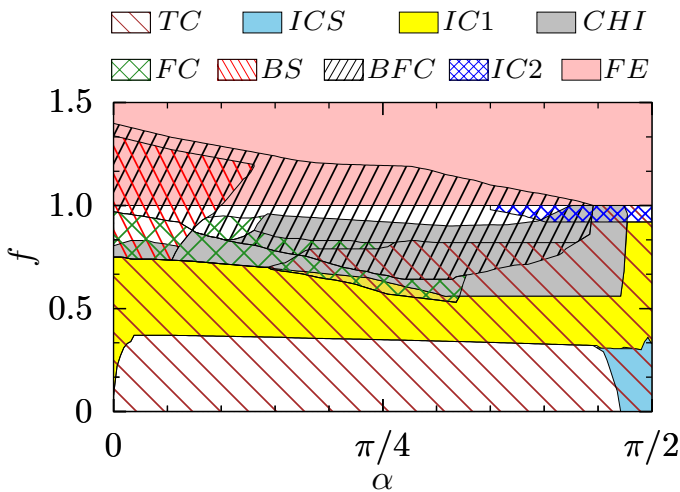}
	\caption{Two-parameter phase diagram in the ($\alpha$, $f$) parameter space depicting various multistable states of the adaptive 
	 network with external periodic forcing. The dynamical states are the same as in Fig.~\ref{fig8},  illustrating a wealth of multistability
	 among the distinct dynamical states for $10$ different initial conditions.
		}
	\label{fig10}
\end{figure}

The multistability between the observed dynamical states is depicted in the two-parameter phase diagram in the ($\alpha$, $f$) parameter space in Fig.~\ref{fig10} for 10 different initial conditions. The dynamical states are the same as in Fig.~\ref{fig8}. Distinct dynamical states are represented as shades and patterns as shown in  Fig.~\ref{fig10}.
 It is evident from the figure that the adaptive network  with external periodic forcing exhibits rich multistability.

\section{Conclusion}
Using the order parameters $R_1$ and $R_2$ along with the space-time plots, snapshots of instantaneous phases, 
and mean frequencies of the collective dynamical states of an adaptive network under the influence of an external periodic forcing,
we have unraveled exotic types of chimera state, namely, itinerant chimeras including bump states and bump frequency cluster states.  
The population density of the oscillators constituting the coherent and incoherent groups of the itinerant chimera co-evolves along with the connection weights, which in turn co-evolve locally, and contribute to the diversity of the frequency distribution of the resulting collective dynamical state.  In addition, the adaptive network is also found to admit  synchronization patterns such as phase and frequency clusters, forced entrained state including the incoherent nature of the phase dynamics. We have also defined two versions of the strength of incoherence to quantify and corroborate the distinct dynamical states and their transitions. An analytical stability condition for the forced entrained state is also found to agree well with the simulation results for appropriate initial conditions of the coupling weight matrix. In view of the  recent profound relevance of chimera states in complex neuronal patterns~\cite{rmjw2010,lmmays1977,mtjwb2016,pjuws2006,smbkb2019,es2020},
we speculate that the observed chimera states may provide clues about the underlying dynamical mechanism of several neuronal and neuropathological states because of their dynamic frequency distribution depending on the co-evolving connection weights and phases.  
Further, as a major motivation for the employed model hails from neuroscience, it is important 
to investigate  the emergence of the observed dynamical patterns in relevant biological models and to explore their
role in the self-organization of adaptive dynamical patterns of neuronal assemblies. 

\begin{acknowledgments}
The work of V.K.C. is supported by the SERB-DST-MATRICS Grant No. MTR/2018/000676 and DST-CRG Project under Grant No. CRG/2020/004353. 
The work of V. K. C. is also supported by DST, New Delhi for computational facilities under the DST-FIST programme (SR/FST/PS-1/2020/135)to the Department of Physics. RB and ES acknowledge support by DFG (Project No. 440145547).  The work of MS forms a part of a research project sponsored by Council of Scientific and Industrial Research (CSIR) under the Grant No. 03(1397)/17/EMR-II.  DVS  is supported by the DST-SERB-CRG Project under Grant No. CRG/2021/000816.
\end{acknowledgments}
 
  \appendix

\renewcommand{\thefigure}{A\arabic{figure}}

\setcounter{figure}{0}

\section*{Appendix  A:   Evolution of coupling weights} 
\begin{figure*}[ht!]
			\includegraphics[width=1.00\textwidth]{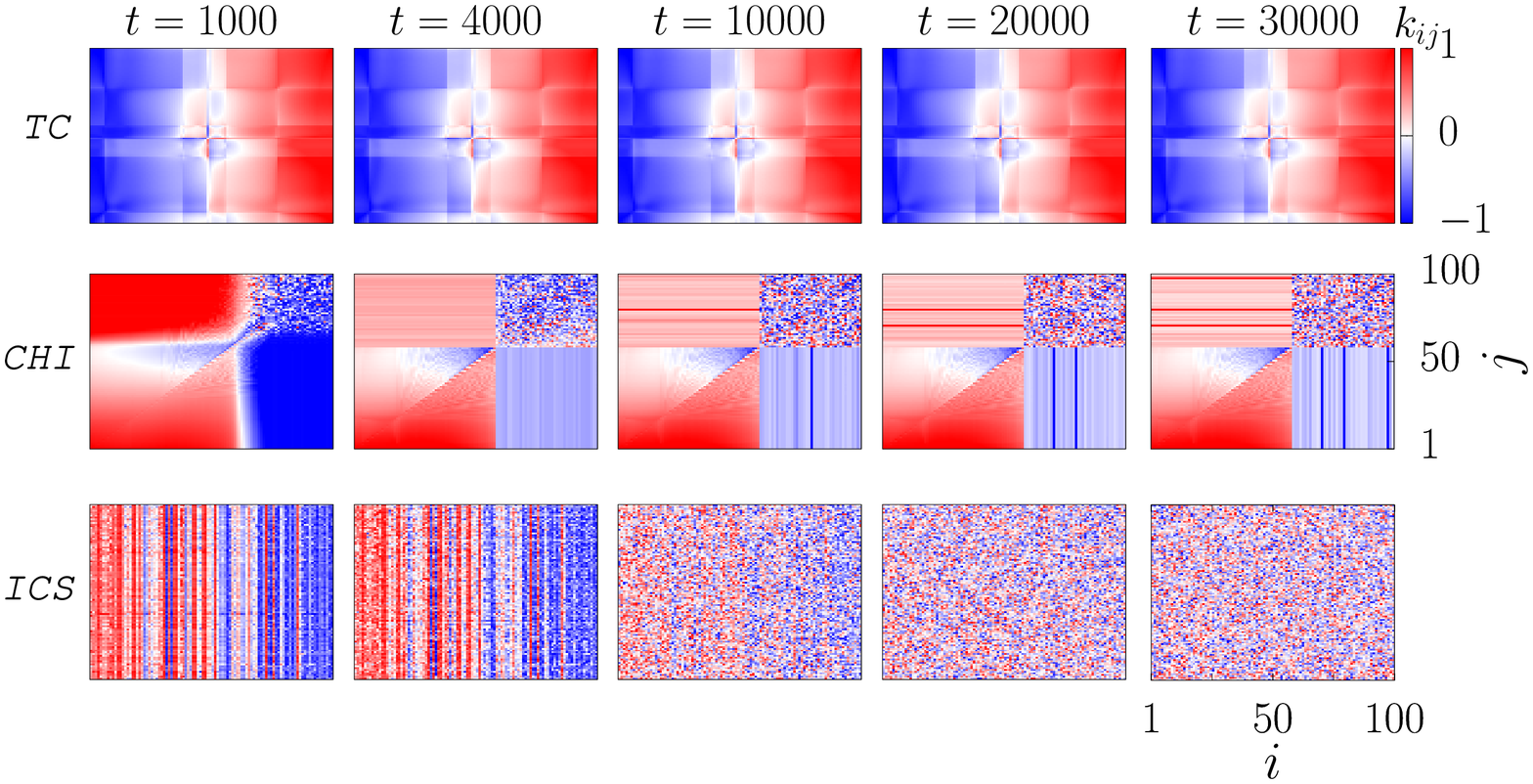}
			\caption{Coupling weight matrix  at  different times. First row represents the two-cluster state for $\alpha=0.40$ and $f=0.00$, second row represents the chimera state $\alpha=0.40$ and  $f=0.78$ and third row represents the incoherent state for $\alpha=1.54$ and $f=0.20$.}
		\label{app.fig1}
	\end{figure*}

\begin{figure*}[ht!]
			\includegraphics[width=1.00\textwidth]{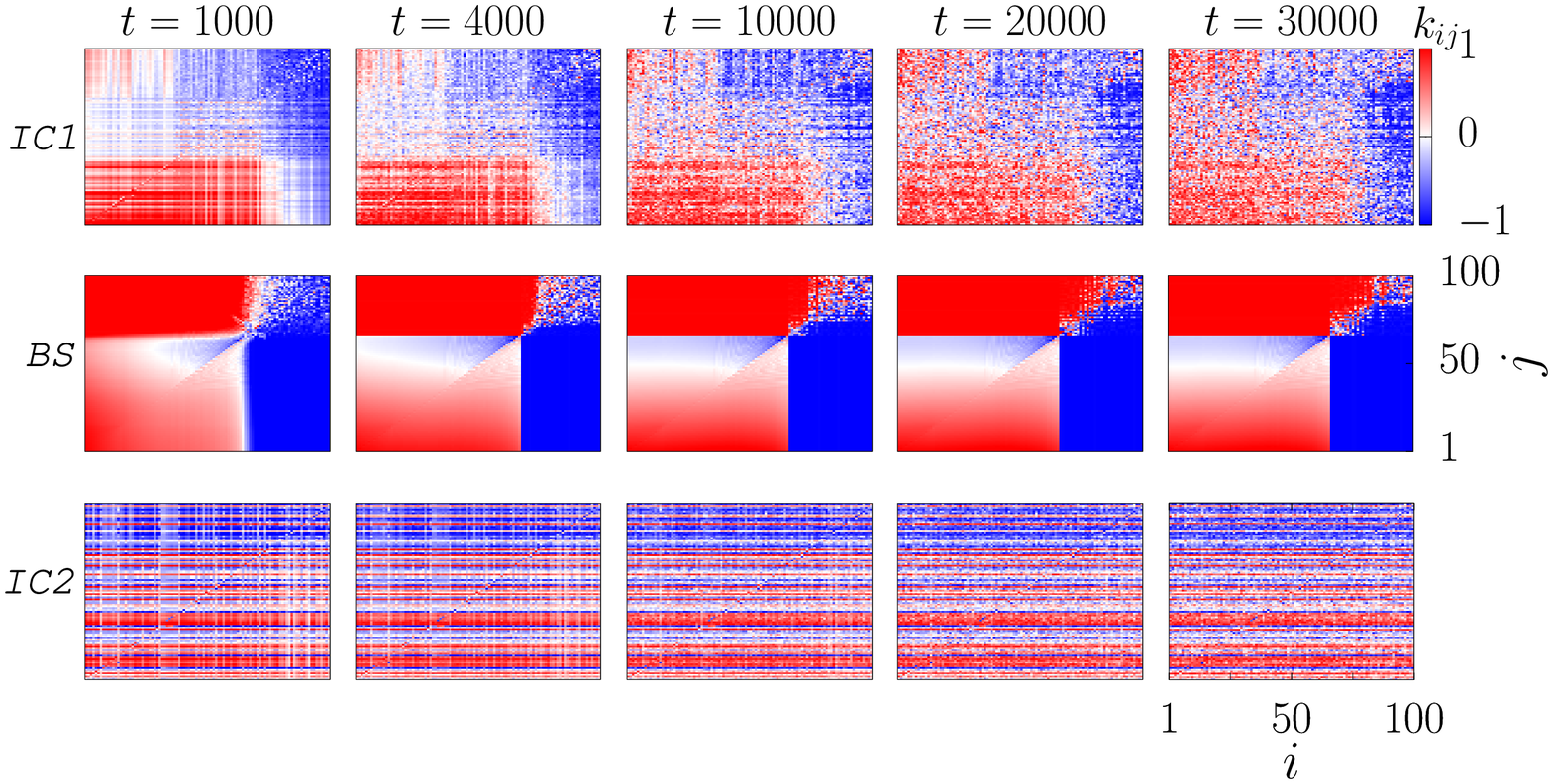}
			\caption{Coupling weight matrix at  different times. First row represents the itinerant chimera of the first kind for $\alpha=1.54$ and  $f=0.78$, second row represents the bump state $\alpha=0.10$ and  $f=0.90$ and third row represents the itinerant  chimera of  the second kind for $\alpha=1.54$ and  $f=0.99$.}
		\label{app.fig2}
	\end{figure*}

\begin{figure*}[ht!]
			\includegraphics[width=1.00\textwidth]{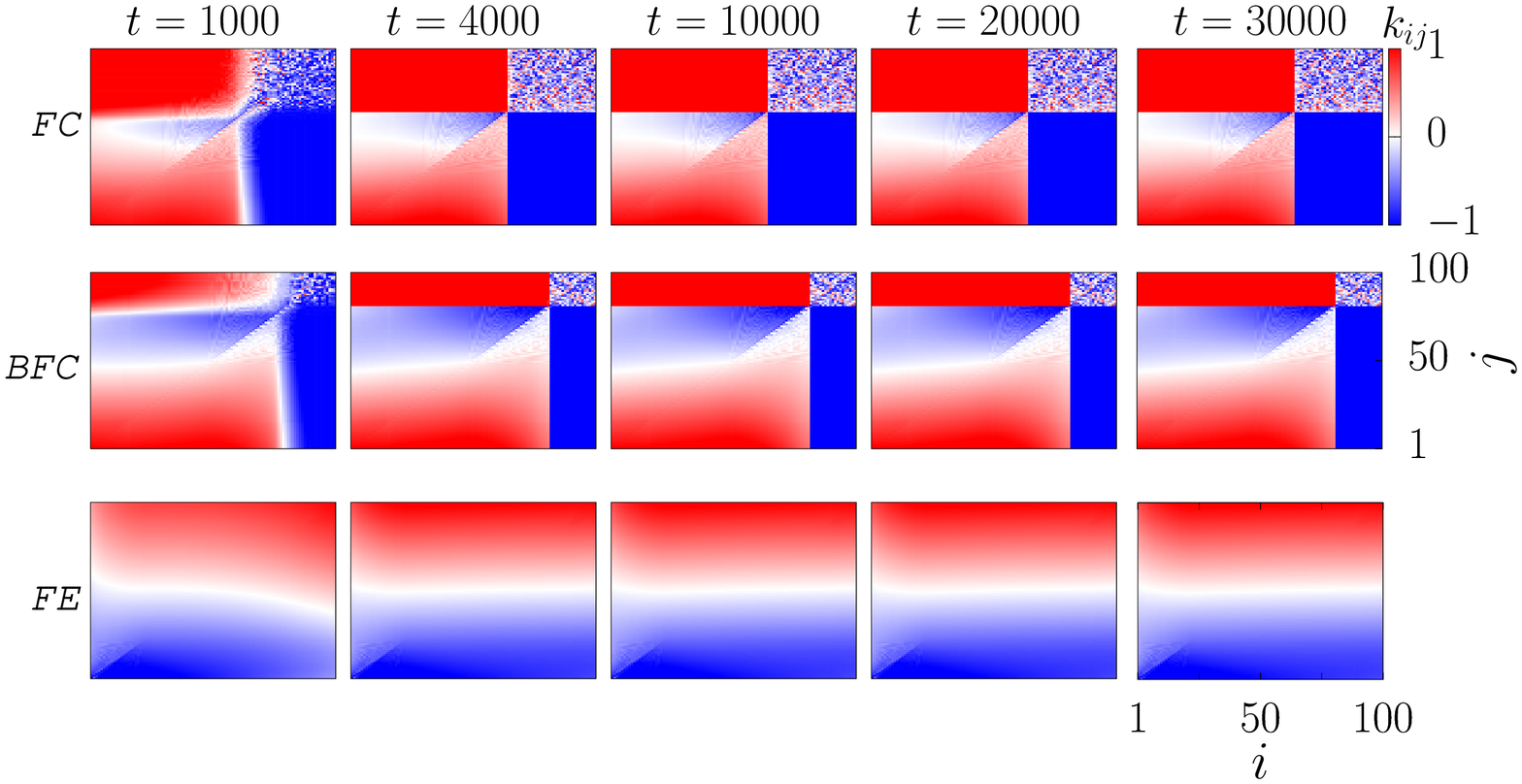}
			\caption{Coupling weight matrix  at  different times. First row represents the frequency cluster state for $\alpha=0.40$ and  $f=0.83$, second row represents 
			bump frequency cluster state   $\alpha=0.60$ and  $f=0.95$  and third row represents the forced entrainment state for $\alpha=0.40$ and  $f=1.40$.}
		\label{app.fig3}
	\end{figure*}

Coupling weights of distinct dynamical states at different time instance are depicted in Figs.~\ref{app.fig1} - ~\ref{app.fig3}.   
Evolution of coupling  weights for two cluster, chimera and incoherent states are depicted in  Fig.~\ref{app.fig1}.
Evolution of coupling  weights for itinerant chimera of  the first kind, bump state and itinerant chimera of the second kind are shown in Fig.~\ref{app.fig2}, while that of frequency cluster, 
bump frequency cluster state and forced entrained state are illustrated in Figs~\ref{app.fig3}. Note that except for the entrained states  such as two-cluster state, frequency cluster and 
frequency entrainment state (where the structure of the coupling matrix does not change qualitatively),  the inter-cluster coupling weights of all other observed states evolve in time.


 \section*{Appendix  B:   Estimation of  the strength of incoherence} 
\begin{figure*}[!ht]
	\includegraphics[width=1.0\textwidth]{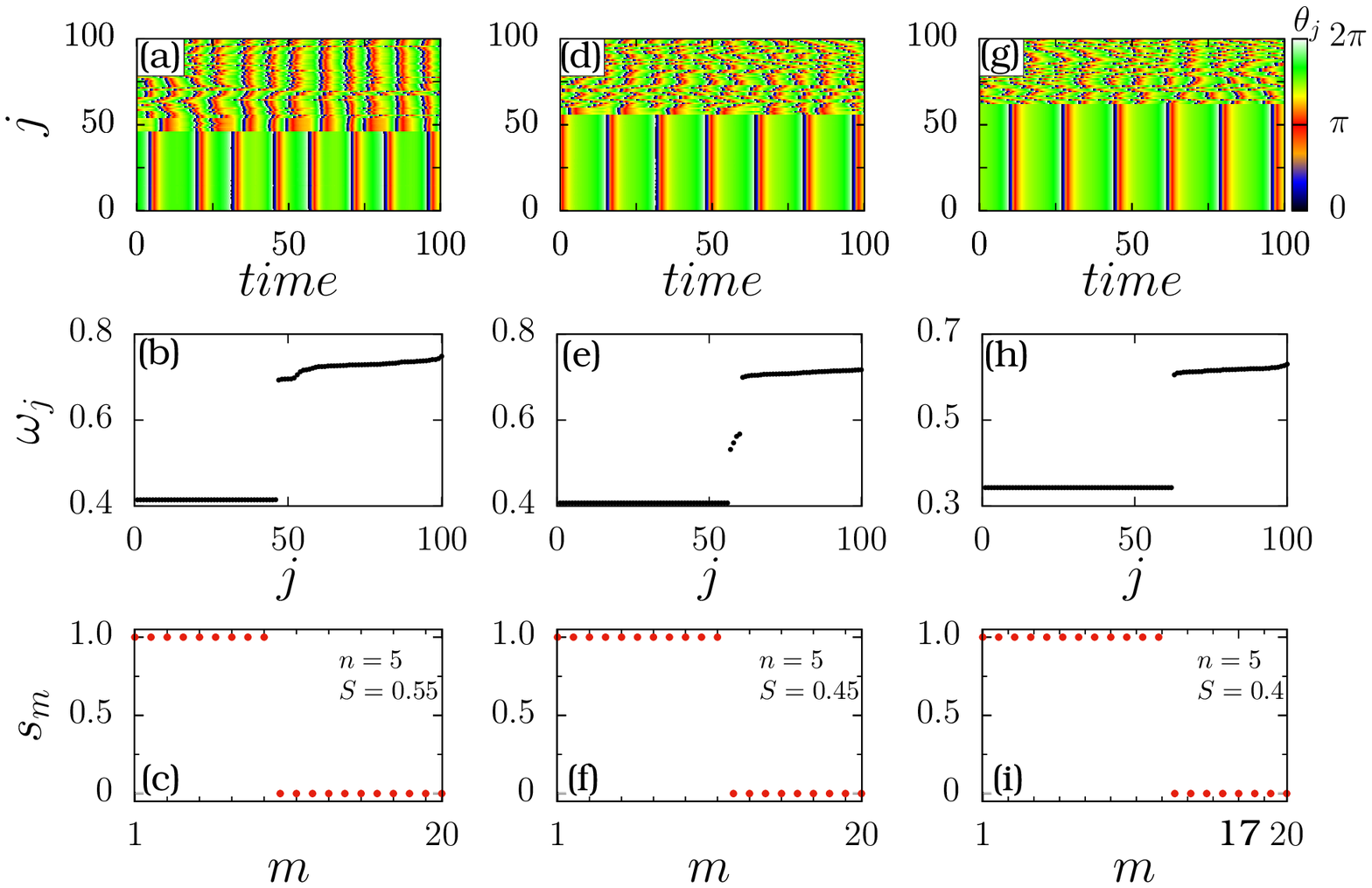}
	\caption{Chimera states for different $f$. First row depicts the time evolution of phases in the range $t\in(26400, 26500)$.  Second
	and third row depicts the time averaged frequencies in the range $t\in(24000, 26500)$ and the strength of incoherence, respectively.
	(a)-(c) $f=0.73$, (d)-(f) $f=0.77$ and (g)-(i) $f=0.81$.  Other parameter values are fixed as $\alpha=0.4$ and $\varepsilon=0.005$.}
	\label{app.fig4}
\end{figure*}

Spatiotemporal plots of the chimera state  for  three different values of the strength of the external forcing $f=0.73, 0.77$ and $0.81$ are depicted
in the first row of  Fig.~\ref{app.fig4}. Other parameter  are fixed as $\alpha=0.4$ and $\varepsilon=0.005$.  Time averaged frequencies of the oscillators,
averaged over the time interval $t\in(24000, 26500)$, are depicted in the second row of  Fig.~\ref{app.fig4}. Both the spatiotemporal plot
and the time averaged frequencies clearly illustrate the co-existing coherent and incoherent  groups of the chimera states.  Now, $N=100$ oscillators,
indexed in the same order  as in  the spatiotemporal and the time averaged frequencies plots, are divided into $m=20$ bins with $n=5$ oscillators in 
each bin.  The standard deviation $\sigma_m$  for each bin is calculated using (4).  $s_m=\Theta(\delta-\sigma_m)$  in each bin 
depicted in the third row of  Fig.~\ref{app.fig4}.  Since,   $s_m=1$ for  the coherent groups and $s_m=0$ for incoherent groups and consequently,
the strength of incoherence take values $S=0$ and $1$ for coherent and incoherent groups, respectively,
which is estimated using (\ref{si}).   The strength of incoherence for the  chimera states in  Fig.~\ref{app.fig4} are also included in the bottom row.

\section*{Appendix  C:  Chimeras for smaller $N$} 
\begin{figure}[!ht]
	\includegraphics[width=0.5\textwidth]{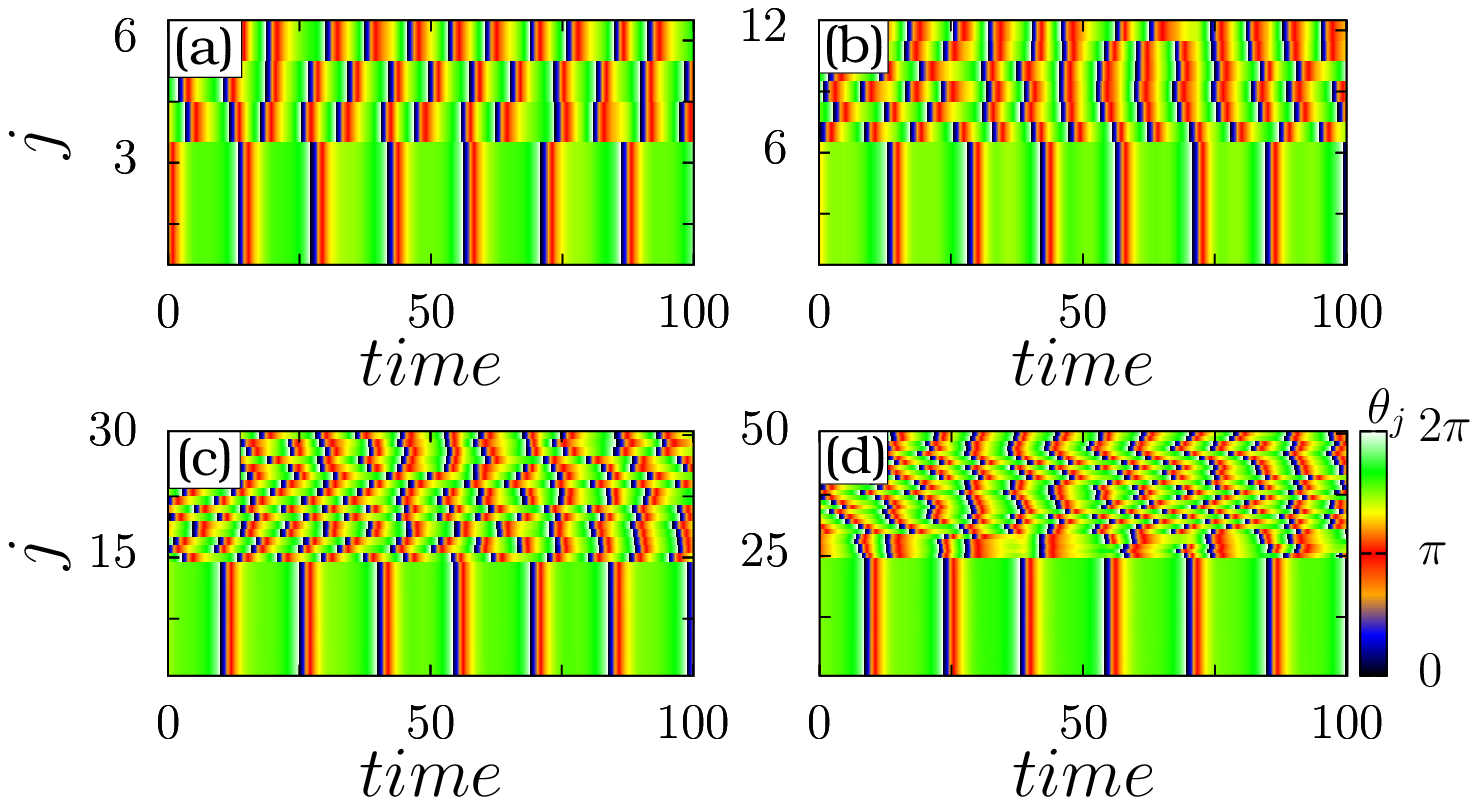}
	\caption{Chimera states for different $N$. (a) $N=6$, (b) $N=12$, (c) $N=30$ and (d) $N=50$.  Other parameter values are fixed
	as $\alpha=0.4, f=0.73$ and $\varepsilon=0.005$.}
		\label{app.fig5}
\end{figure}

Chimera states for $N=6, 12, 30$ and $50$ are depicted in Fig.~\ref{app.fig5}(a)-~\ref{app.fig5}(d), respectively, to illustrate the emergence of chimera
states even for smaller $N$.  Other parameter values are fixed
as $\alpha=0.4, f=0.73$ and $\varepsilon=0.005$. 

\begin{thebibliography}{99}
\bibitem{note} Note that one needs to be very careful about the choice of $M$.
\end{thebibliography}
\end{document}